\documentclass[%
 reprint,
 superscriptaddress,
 amsmath,amssymb,
 aps,
 prl,
]{revtex4-2}

\usepackage{graphicx}
\usepackage{dcolumn}
\usepackage{bm}
\usepackage{color}
\usepackage{slashed}
\usepackage[colorlinks,allcolors=cyan]{hyperref}
\usepackage{physics}
\usepackage[caption=false]{subfig}
\usepackage{ragged2e}
\DeclareCaptionOption{justified}[]{\caption@setjustification{justified}}
\usepackage[capitalise]{cleveref}
\newcommand{\Zt}{\mathbb{Z}_2}
\newcommand{\rr}{{\vb{r}}}

\newcommand{\beq}{\begin{equation}}
\newcommand{\eeq}{\end{equation}}

\begin{document}

\title{Deconfined quantum criticality in Ising gauge theory entangled with single-component fermions}

 \author{Umberto Borla}
 \affiliation{Racah Institute of Physics, The Hebrew University of Jerusalem, Jerusalem 91904, Israel}
 \author{Snir Gazit}
 \affiliation{Racah Institute of Physics, The Hebrew University of Jerusalem, Jerusalem 91904, Israel}
 \affiliation{The Fritz Haber Research Center for Molecular Dynamics,
The Hebrew University of Jerusalem, Jerusalem 91904, Israel}
\author{Sergej Moroz}
 \affiliation{ Department of Engineering and Physics, Karlstad University, Karlstad, Sweden}
 \affiliation{Nordita, KTH Royal Institute of Technology and Stockholm University, Stockholm, Sweden}

\begin{abstract}
 We highlight exotic quantum criticality of quasi-two-dimensional single-component fermions at half-filling that are minimally coupled to a dynamical Ising gauge theory. With the numerical matrix product state based iDMRG method, we discover a robust quantum critical line in the infinite cylinder geometry, where gauge confinement and dimerized translation symmetry breaking emerge simultaneously. We investigate how the transition can be split by a $\mathbb{Z}_2$ topologically ordered dimerized phase that is stabilized by additional short-range repulsive interactions. We conjecture a $u(1)$ deconfined criticality scenario, propose a corresponding low-energy effective field theory of the exotic quantum critical point in the two-dimensional limit and identify its shortcomings.
\end{abstract}

\maketitle

\emph{Introduction --- } Investigating quantum critical phenomena that defy an interpretation in terms of the Landau theory of spontaneous symmetry breaking (SSB) is central to modern condensed matter theory \cite{Fradkin2013,wenbook, Sachdevbook, sachdev_2023}. A case in point is provided by confinement and Higgs transitions, captured by the condensation of magnetic and electric particles respectively, that are not accompanied by spontaneous symmetry breaking of ordinary global symmetries \cite{Fradkin_1979}. Much of the ongoing research is aimed at generalizing the concept of symmetry \cite{gaiotto2015generalized} and extending the Landau paradigm to fractionalized quantum phases of matter and associated phase transitions \cite{mcgreevy2023generalized}.
\begin{figure}[t]
\captionsetup[subfigure]{labelformat=empty}
    \subfloat[\label{subfig:model_a}]{}
    \subfloat[\label{subfig:model_b}]{}
    \subfloat[\label{subfig:phase_diag}]{}
    \hspace{-10pt}
	\includegraphics[width=1.0\linewidth]{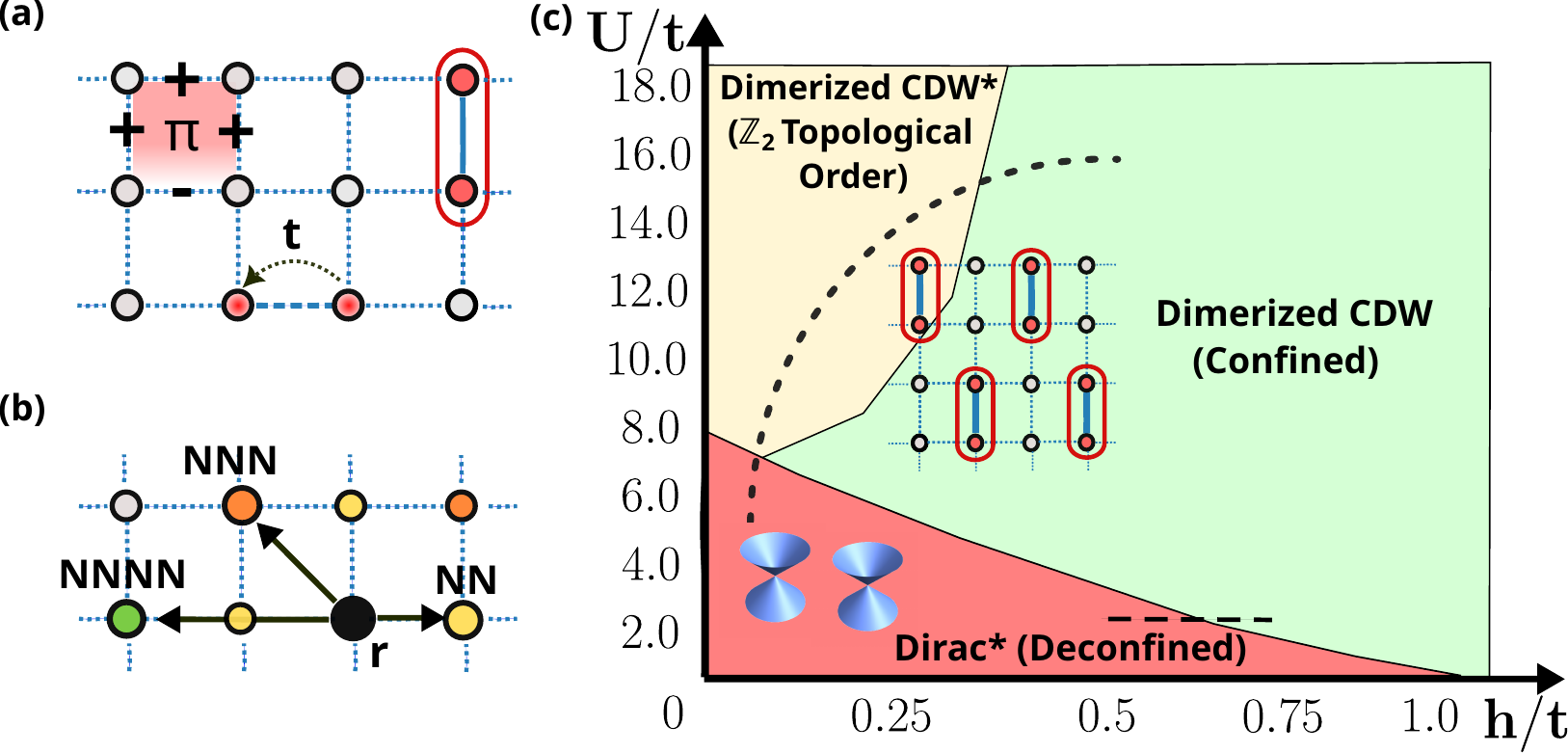}
	\caption{(a) The model consists of single-component fermions hopping between sites of a square lattice and $\Zt$ gauge fields living on links. The Hilbert space is formed by fermionic pairs connected by an ``electric" string (solid blue line). In the deconfined regime, a background $\pi$-flux appears. In (b) we highlight the nature of the repulsive density-density interactions, which stabilize the D-CDW* phase. Given a site at $\rr$, we illustrate the nearest neighbors (yellow), next-nearest neighbors (orange) and next-next-nearest neighbors (green) that enter the definition \eqref{eq:U_int}. (c) Quantum phase diagram as a function of $h/t$ and $U/t$, based on the numerical results obtained for a cylinder of circumference $L_y=4$. The ratio $J/U=-1/40$ is fixed to stabilize the dimerized topologically ordered phase at large $U$, as explained in the main text and \cite{SM}. The horizontal dashed line corresponds to $U/t=2$, as scanned in Fig. \ref{subfig:O_single_trans_finite_U}. The curved dashed line corresponds to the cut in parameter space scanned in Fig. \ref{subfig:O_split_trans}, which is parametrized as $(h/t,U/t)=(0.3+0.25 \cos \lambda,4+12 \sin \lambda)$.}
    \label{fig:pd}
\end{figure}

A natural framework for realizing such exotic transitions is lattice gauge theories (LGTs), often used to study strong interactions between fluctuating gauge and matter degrees of freedom \cite{wegner1971duality, Wilson1974, Kogut75,Kogut1979}.
In the context of condensed matter physics, LGTs emerge as an effective low-energy description, comprising objects which are fractionalized with respect to the underlying symmetry and interact via gauge field fluctuations. This setting arises from microscopic models that describe, e.g., spin liquids \cite{savary2016quantum} and fractional quantum Hall states \cite{wilczek1990fractional}. In a parallel effort, recent breakthrough progress in quantum engineering is anticipated to provide a complementary "bottom-up" approach,  where lattice gauge theories can be simulated reliably in a controlled artificial environment \cite{Dalmonte16,banuls2019simulating, banuls2020review,halimeh2023coldatom}. 

A particularly fertile ground is provided by models that exhibit an interplay between fermionic matter and gauge fields, see e.g. \cite{Senthil_2000, Hermele_2004, Assaad_2016, Gazit_2017, xu2019, Chen_2020, wang2019, Gazit2020, janssen2020}. In this context, recent work \cite{Gazit_2018} has put forward a critical Higgs theory where the above two paradigms, symmetry breaking and confinement, occur in tandem. More specifically, quantum Monte Carlo simulations of spinful lattice fermions minimally coupled to an Ising gauge theory revealed a single and continuous deconfined criticality, where the appearance of antiferromagnetic order coincides with confinement. Generalizations of this exotic scenario, particularly to models with a different number of fermion flavors, remain an open inquiry. 

In this letter, we employ numerical iDMRG simulations on infinite cylinders to investigate the confinement transition of single-component lattice fermions coupled to a fluctuating Ising gauge theory. At half-filling we find that increasing the gauge field electric string tension nucleates a continuous transition, where translational symmetry breaking, in the form of fermion dimer VBS states, precisely coincides with the confinement of matter fields. Crucially, we provide numerical evidence for robustness of the transition by testing its stability with respect to additional weak short-ranged density-density interactions. By contrast, for sufficiently strong interactions,  confinement and SSB transitions split giving rise to a $\mathbb{Z}_2$ topologically ordered dimerized charge density wave phase. We suggest and scrutinize a field theory description of the exotic transition in terms of a double-charged Higgs field condensation in a compact $u(1)$ gauge theory coupled to a pair ($N_f=2$) of two-component Dirac fermions. 

\emph{Model and phase diagram} --- As a concrete lattice model, we study single-component fermions coupled to $\Zt$ gauge fields, see Fig. \ref{subfig:model_a}. The gauge degrees of freedom are Ising spins $\sigma^z_b$ residing on the square lattice bonds $b=\{\rr,\eta\}$, with site labels $\mathbf{r}$ and unit vectors $\eta=\hat{x},\hat{y}$. Complex single-component fermions that carry gauge charge are created by the raising operators $c_\rr^\dagger$. The quantum dynamics is governed by the Hamiltonian $H=H_{\Zt}+H_f$. The first term is the standard Ising lattice gauge theory \cite{Kogut75}
\begin{equation}
H_{\Zt} = -J \sum_{\square} \prod_{b\in \square} \sigma^z_b -h \sum_{\mathbf{r},\eta} \sigma^x_{\rr,\eta},
    \label{eq:H_z2}
\end{equation}
where $\sigma^{\gamma}_b$ for $\gamma=\{x,y,z\}$ are Pauli operators. 
The second term of the Hamiltonian,
\begin{equation}
    H_f= -t\sum_{ \mathbf{r},\eta} \left( c^\dagger_{\mathbf{r}}\sigma^z_{\mathbf{r},\eta} c_{\mathbf{r}+\eta} + \text{h.c.} \right) 
 -\mu\sum_{\mathbf{r}}c^\dagger_{\mathbf{r}}c_{\mathbf{r}}
\end{equation}
incorporates hopping events of fermions minimally coupled to the $\Zt$ gauge fields. In addition, we include a finite chemical potential in order to fine-tune to half-filling. In what follows, we measure all microscopic energy scales in units of the hopping amplitude $t$.

The Hamiltonian admits a global $U(1)$ symmetry associated with fermion number conservation and respects the square lattice translations and point group symmetries. The gauge structure is revealed via the invariance with respect to local transformations generated by the operators 
\begin{equation}
    G_\rr=\prod_{\eta \in +_\rr}\sigma^x_{\rr, \eta} (-1)^{n^f_\rr},
\end{equation}
where the product is taken over the four links emanating from the site, which obey $\qty[H,G_\rr]=0$ for all sites $\rr$. Importantly, we choose to work in a specific sector obeying the ``even'' Gauss's law $G_\rr=+1$, corresponding to vanishing background $\Zt$ charges.

In the large $h$ regime, the ``electric'' term in \eqref{eq:H_z2} dominates, leading to charge confinement. Together with the Pauli exclusion principle, this forces fermion pairs to combine into dimers residing on nearest-neighbor sites and obeying a hard-core constraint. At half filling, the effective repulsive dimer-dimer interaction stabilizes a staggered dimer charge density wave (D-CDW) pattern that spontaneously breaks translation and point group symmetries ~\cite{borla2022}, see \cref{subfig:phase_diag}. On the other hand, in the opposite limit, $h\rightarrow 0$, the gauge theory is deconfined, allowing fermionic matter to propagate. Following Lieb's theorem \cite{lieb1994flux,Affleck_1988}, the background magnetic Ising flux is fixed to a $\pi$-flux pattern. The resulting dispersion hosts two Dirac cones, which is a neutral semimetal at half-filling. Therefore in this phase, while magnetic excitations are gapped, due to the gapless Dirac spectrum one anticipates power-law correlations of the gauge-invariant fermion Green's function, where fermions are connected by a string of $\mathbb{Z}_2$ gauge fields.

As the two limiting cases are distinct phases of matter, they must be separated by at least one quantum phase transition. A previous numerical analysis~\cite{borla2022} suggests a single quantum critical point (QCP) at $h_c$ where D-CDW order and confinement coincide. This result is surprising since, without fine-tuning of microscopic parameters, either a first order or split phase transition is expected. For the latter, an intermediate phase, CDW$^*$, can form, characterized by SSB and topological order \cite{Senthil_2000,Gazit_2018}. Understanding the physical nature of the exotic, direct, and continuous transition is an outstanding problem that forms the basis of this study.

To address this query, we first seek to individually tune the appearance of SSB and confinement and enable a split phase transition. We achieve this goal by turning on a specific combination of extended short-range density-density repulsive interactions controlled by a single coupling $U$. Explicitly, we add to the Hamiltonian a term of the form  
\begin{equation}
H_U=U\qty[\sum_{\langle \rr,\rr'\rangle}n_\rr n_{\rr'}+\sum_{\langle \langle \rr,\rr' \rangle \rangle}\frac{n_\rr n_{\rr'}}{2}+\sum_{\langle \langle \langle \rr,\rr' \rangle \rangle \rangle}\frac{n_\rr n_{\rr'}}{4}],
\label{eq:U_int}
\end{equation} 
see \cref{subfig:model_b}.
As detailed in \cite{SM}, in the limit $U \rightarrow \infty$ this particular choice of coefficients forces the system into a classical ground state with an identical symmetry breaking pattern as the large $h$ regime of the gauge theory. In the following, using numerically exact methods, we test this scenario by elucidating the global zero temperature phase diagram of the generalized model at half-filling as a function of $h/t$ and $U/t$, see \cref{subfig:phase_diag}.
    
\emph{Numerical methods and observables} --- The gauge constraint $G_\rr=1$ introduces a numerical sign problem, and hence the quantum Monte Carlo methods introduced in Refs.~\cite{Assaad_2016,Gazit_2017} are inapplicable for the single-component case. To address the problem numerically, we employ a variational matrix product state (MPS) approach, which we optimize via the DMRG algorithm \cite{PhysRevLett.69.2863, mcculloch2008infinite, hauschild2018efficient} on infinite (along the $x$ axis) cylinders of circumference $L_y=4$. While in two spatial dimensions the MPS representation suffers from an exponential scaling of the computational complexity as a function of $L_y$, valuable information on the true (quasi-two-dimensional) thermodynamic behavior can be inferred by studying finite-circumference cylinders.

To characterize the formation of the SSB dimerized CDW pattern, we consider the order parameter $\mathcal{O}^{\eta}_{\text{D-CDW}}= n_{max}-n_{min}$, which measures the particle number imbalance within a unit cell. This by itself does not provide information on the specific arrangement of the charges, which is however easily obtained by direct inspection of the local fermionic density.
While on the infinite plane the CDW ground state is eightfold degenerate and generated via $\pi/2$ rotations and translations by primitive vectors, in the infinite cylinder geometry the degeneracy is lifted to four-fold.

Detecting confinement in the presence of matter fields is subtle since the standard Wilson loop approach fails due to charge screening. To overcome this difficulty, we compute the magnetic Fredenhagen-Marcu string order parameter \cite{Fredenhagen_1986,Gregor_2011,Verresen_2021}, which probes confinement by detecting the condensation of $m$ particles (visons). This is defined by the ratio $\mathcal{O}_{\text{FM}}^m=\langle \tilde{W}_\text{half}\rangle/\sqrt{\langle \tilde{W}_\text{full}}\rangle,$
where $\tilde{W}_{\text{full}}$ is a square 't Hooft loop and $\tilde{W}_{\text{half}}$ is the open line obtained by cutting the full loop in two, see \cite{SM}. This normalization ensures that the order parameter takes a finite value in the confined phase.

Lastly, when employing iDMRG, the correlation length $\xi$ corresponding to the smallest excitation gap in the system can be extracted from the MPS transfer matrix \cite{hauschild2018efficient, SM}. A divergence of $\xi$ with bond dimension provides a signature of quantum criticality independent of the specific nature of the transition.

\begin{figure}[t]
    \captionsetup[subfigure]{labelformat=empty}
    \subfloat[\label{subfig:O_single_trans}]{}

    \subfloat[\label{subfig:O_single_trans_finite_U}]{}
    \subfloat[\label{subfig:O_split_trans}]{}
    \hspace{-10pt}
    \includegraphics[width=\linewidth]{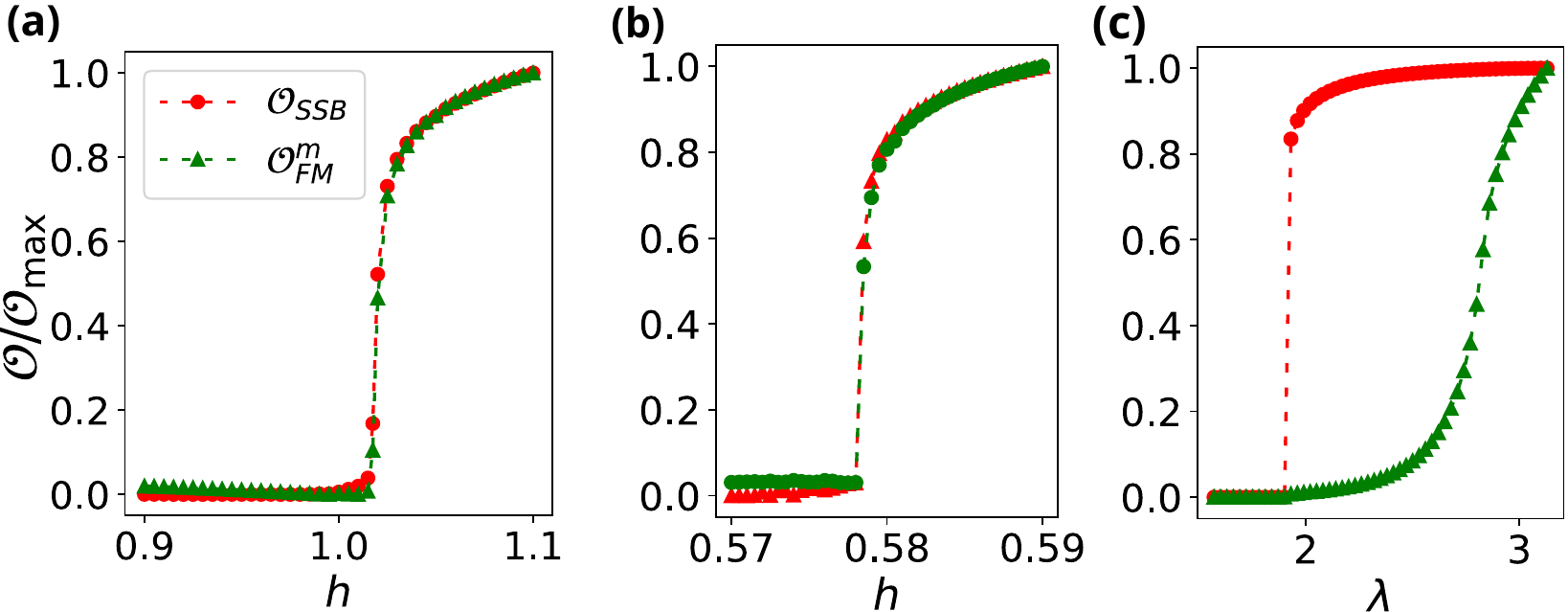}
    \vspace{-10pt}
	\caption{Normalized order parameters for the translation symmetry breaking (red circles) and confinement (green triangles) transitions along the horizontal cuts $U=0$ (left), $U=2$ (center) and along the curve shown in Fig. \ref{subfig:phase_diag} (right). Along the former two a single transition is detected, suggesting the presence of exotic quantum criticality. Along the latter, on the other hand, two transitions are detected, with symmetry breaking occurring before confinement.}
	\label{fig:cut}
\end{figure}

\emph{Numerical results} --- We carry out iDMRG simulations \cite{hauschild2018efficient} for $L_y=4$, monitoring the ground state convergence by considering an increasing range of bond dimensions up to $\chi=1200$ \cite{SM}. For computational convenience, we use the dual spin 1/2 model, introduced in \cite{borla2022} and reviewed in \cite{SM}.

We note that the presence of a negative magnetic term is crucial to stabilize a deconfined phase with a $\pi$-flux background. While for small values of $U$, such a term is perturbatively generated by fermionic hopping, at large $U$ this is no longer the case since the fermions are essentially frozen in the dimerized pattern. For this reason, we choose a negative magnetic coupling $J$ that grows linearly with $U$ \cite{SM}. 

\begin{figure}[t]
\captionsetup[subfigure]{labelformat=empty}
    \subfloat[\label{subfig:corr_single_U0}]{}
    \subfloat[\label{subfig:corr_single_U2}]{}
    \subfloat[\label{subfig:corr_double}]{}
    \hspace{-10pt}
    \vspace{-10pt}
	\includegraphics[width=0.9\linewidth]{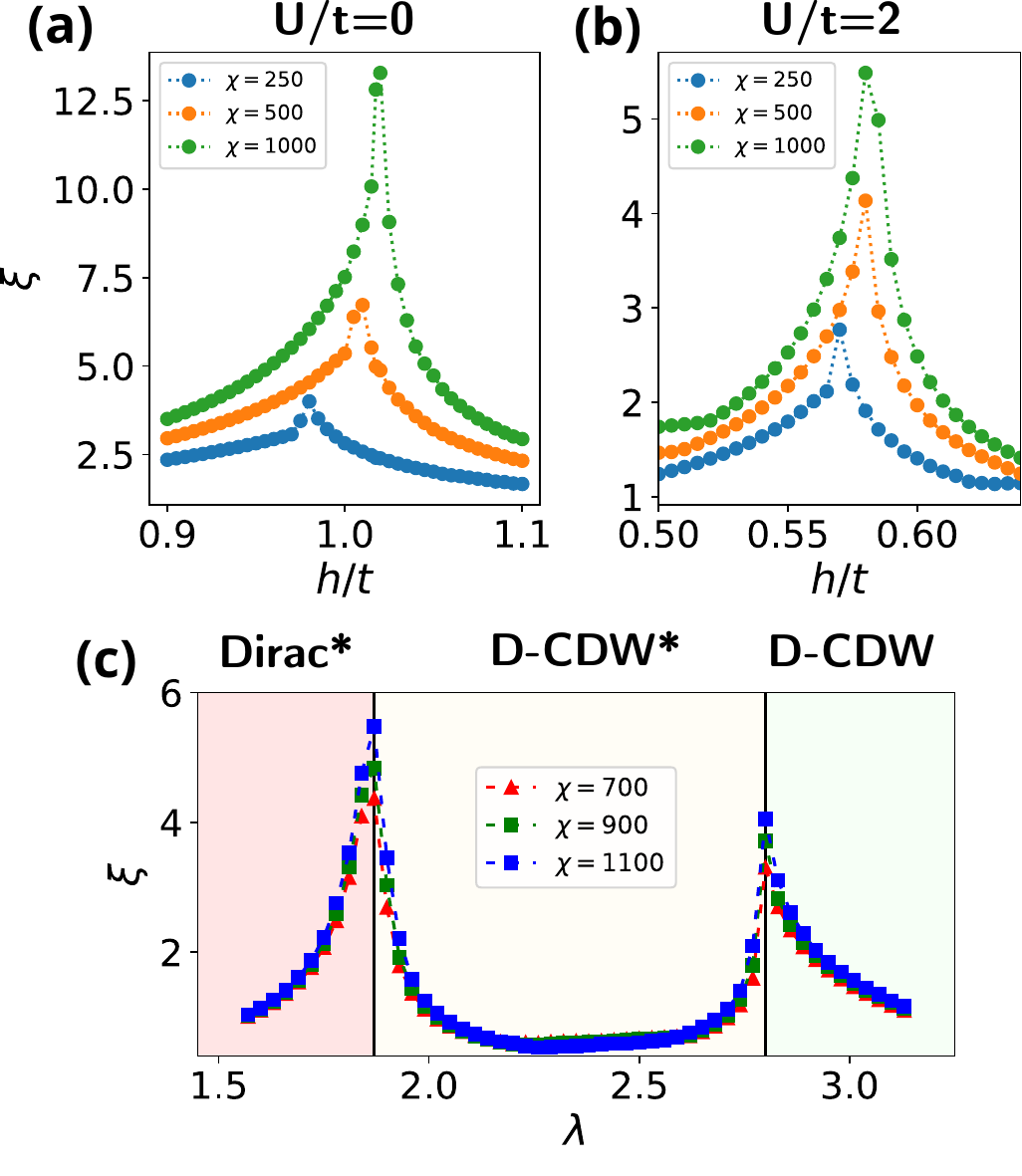}
	\caption{Correlation length $\xi$ extracted from the iMPS transfer matrix along the the horizontal lines $U=0$ (a), $U=2$ (b) and along the curve shown in Fig. \ref{subfig:phase_diag} (c). Peaks in the correlation length, which sharpen as the bond dimension $\chi$ is increased, provide evidence for continuous phase transitions. The peaks coincide, respectively, with the points where the SSB and FM order parameters start to grow, as shown in Fig. \ref{subfig:O_split_trans}.}
	\label{fig:corrs}
\end{figure}

Our main results are summarized in \cref{subfig:phase_diag}. At $U=0$ a single quantum phase transition characterized by simultaneous confinement and translational symmetry breaking occurs. This is evident from \cref{subfig:O_single_trans}, where both order parameters raise simultaneously above the critical field $h_c/t\approx1.01(1)$. To test the stability of the transition, we now drive confinement along a parameter cut $U/t=2$. Indeed, in \cref{subfig:O_single_trans_finite_U}, we observe the same qualitative behavior as for $U/t=0$, displaying a  coincident growth of both order parameters at the critical point $h_c/t\approx0.578(2)$.  This key observation shows that the exotic critical point does \textit{not} require fine-tuning, as it is realized along an extended range of parameters. To provide additional numerical evidence for the continuous nature of the transition, in \cref{subfig:corr_single_U0,subfig:corr_single_U2} we plot the correlation length $\xi$ for increasing values of the bond dimension $\chi$, in proximity to the critical point, for $U/t=0$ and $U/t=2$, respectively. We observe a clear sharpening of $\xi$ supporting the continuous transition conjecture. The above features render it a candidate for a new type of quantum criticality. 

For sufficiently large $U$, the above scenario must break down. More specifically, in the limit $h\to0$, due to \eqref{eq:U_int} we expect an instability of the Dirac$^*$ semimetal towards a D-CDW$^*$ phase. This state is characterized by $\mathbb{Z}_2$ topological order together with translation SSB in the form of a dimer crystal. By increasing $h$, one eventually confines the gauge fields nucleating the D-CDW phase. We test the large $U$ split transition scenario, predicted above, by choosing a trajectory parametrized by $\lambda$, see \cref{subfig:phase_diag} that is anticipated to interpolate between these phases. The result of this analysis is shown in Fig. \ref{subfig:O_split_trans}. We observe that by increasing $\lambda$, the D-CDW order parameter associated with SSB, settles in prior to the rise of Fredenhagen-Marcu order parameter that signals confinement.

To further support the split transition scenario, in \cref{subfig:corr_double}, we depict the correlation length $\xi$ as a function of $\lambda$ and observe two peaks that progressively sharpen with the bond dimensions precisely at the respective locations of the SSB and confinement transition. 
We have, therefore, identified three distinct regions, shown in the phase diagram of Fig. \ref{subfig:phase_diag}. Besides the Dirac* and D-CDW phases, the density-density interactions in \cref{eq:U_int} stabilize an intermediate D-CDW* phase, where a dimerized translationally broken symmetry pattern coexists with $\mathbb{Z}_2$ topological order \cite{Senthil_2000}. 

\emph{Towards 2d criticality} --- Motivated by the iDMRG numerical phase diagram, which is inherently restricted to cylindrical geometries, we now discuss extensions of the various putative quantum critical points to infinite two-dimensional lattices.

We first examine the case of frozen gauge fluctuations $h=0$. Here electric strings have vanishing tension and the Ising gauge $\pi$-flux background is static, imposed by a 1-form magnetic symmetry. Due to a mixed t'Hooft anomaly between this symmetry and the particle number $U(1)$, the system cannot be trivially gapped \cite{borla2021gauging, su2024gapless}. Progressively increasing the interaction strength $U$ eventually gaps the Dirac fermions, nucleating the topologically ordered D-CDW$^*$. As shown by a simple mean field analysis \cite{SM}, the zig-zag dimerized pattern of translation symmetry breaking ensures that the order parameter couples to the emergent Dirac fermions as a mass term. This implies a fractionalized Gross-Neveu* scenario \cite{PhysRevLett.125.257202} for the deconfined Dirac* to the  D-CDW* phase transition.

To gain some insight into the nature of the confinement transition at large $U$, we go deep into the translationally broken regime where fermions are fully frozen into the dimerized pattern. These act as static charges, realizing a gapped pure $\mathbb{Z}_2$ gauge theory with a non-uniform Gauss law $G_\rr=\mp1$ on sites with/without fermions. Using Wegner's non-local duality \cite{wegner1971duality}, we can rewrite the Ising gauge theory as a \emph{half-frustrated} Ising model defined on a dual square lattice, see \cite{SM} for details. Using this analysis, we find that across the confinement transition visons condense at zero momentum, similarly to the uniform Ising gauge theory. The resulting confinement transition is known to belong to the Ising* universality class.

Most intriguing, our numerical results indicate that for sufficiently small $U$ the confinement transition coincides with the breaking of lattice translations forming a D-CDW phase. Crucially, as we explicitly demonstrated, this does not seem to involve fine-tuning since, as a function of the coupling $U$, the critical point evolves into a continuous line of critical points. 

Inspired by the spinful case studied in Ref.~\cite{Gazit_2018}, we conjecture a Higgs mechanism that enables driving both transitions via a \textit{single} tuning parameter identified with the Higgs mass. Concretely, we suggest enlarging the $\Zt$ gauge redundancy to a compact $u(1)$ gauge group. In addition, we introduce a complex scalar Higgs field $H$ carrying two units of charge under the emergent $u(1)$ gauge group. When the Higgs field is condensed, the $u(1)$ gauge redundancy is reduced to $\Zt$, such that the resulting phase is a Dirac semimetal coupled to a deconfined $\Zt$ gauge theory. On the other hand, when the Higgs field is massive, the low energy theory is described by $N_f=2$ two-component Dirac fermions $\psi_i$ coupled to a compact gauge field $a_\mu$. As argued by Polyakov \cite{POLYAKOV_1975}, in the pure compact $u(1)$ case, the proliferation of monopoles always leads to confinement. When coupled with $N_f$ flavors of Dirac fermions, it is known that for sufficiently large values of $N_f$ greater than a critical value $N_f^c$, a deconfined phase is realized \cite{Hermele_2004}. Large-$N_f$ calculations \cite{PhysRevD.89.065016} and quantum Monte Carlo simulations \cite{PhysRevD.100.054514} indicate that $N^c_f\approx 12$. Therefore, our case with $N_f=2$ is confined with a VBS order of the resulting dimers. We conjecture the following low-energy effective theory 
\begin{align}
\mathcal{L}=&\sum_{i=1}^2\bar{\psi}_i \gamma_\mu\qty(i\partial_\mu+a_\mu)\psi_i +\mathcal{L}_\mathcal{M}+ \nonumber \\
&\qty|\qty(i\partial_\mu+2a_\mu)H|^2+V(H).
\label{eq:higgs_ft}    
\end{align}
In the above equation, the Higgs potential takes the standard form $V(H)=m |H|^2+U|H|^4$ and $\mathcal{L}_M$ represents contributions arising from symmetry-allowed magnetic monopoles. Exactly at the critical point a gapless photon should appear because  magnetic monopoles
are expected to be RG-irrelevant. This in turn gives rise to an emergent global $U(1)$ symmetry associated with the
magnetic flux conservation of the u(1) gauge field.
Physically this corresponds to a continuous $SO(2)$ spatial rotation, enlarging the physical $C_4$ symmetry that rotates between the four distinct dimer CDW patterns at the lattice scale.

The theory \eqref{eq:higgs_ft} might have a subtle, but important deficiency. Following \cite{borokhov2003topological}, an elementary monopole operator comes with $N_f/2=1$ fermionic zero-mode. Deep in the Higgs phase, where the gauge magnetic flux background is $\pi$,  this fermionic mode transforms projectively under elementary spatial translations resulting in non-commutative transformations in the two orthogonal directions. Provided the $2\pi$ flux insertion transforms linearly, the $u(1)$ monopole (that exist above the Higgs mass scale) should thus also transform projectively. This is inconsistent, however, with our microscopic lattice theory, where elementary translations commute when acting on local gauge-invariant operators. Given that the Higgs scale vanishes at the transition, this ultraviolet issue descends to infrared and thus poses a consistency problem for the effective theory of the critical point.

\emph{Outlook } --- We conclude our presentation by identifying several future extensions of our results: 
First, in the deconfined regime in the $\pi$-flux background two Dirac Fermi surfaces should emerge as we move away from half-filling. Attractive interaction mediated by the Ising electric field should give rise to fermion p-wave pairing whose node structure and chirality is worth investigating.
Second, note that the confinement-induced fermion dimers do not necessarily have to form a crystalline pattern. Alternatively, they can form a superfluid state or even more exotic spin liquid states depending on the effective dimer hopping amplitude, inter-dimer interaction strength, and lattice geometry. It would be interesting to stabilize such phases and explore the fate of the confinement transition in this setting. 
Finally, to scrutinize further the conjectured field theory description of the deconfined transition \cref{eq:higgs_ft}, it is desirable to develop a microscopic lattice construction of the Higgs field and a precise assignment of global symmetries in the spirit of Ref. \cite{Gazit_2018}. This will hopefully clarify the fate of the particle-hole symmetry and how elementary translations act on $u(1)$ monopoles.

\emph{Acknowledgements}--- We thank Lukas Janssen, Ruben Verresen, Subir Sachdev, and Fakher Assaad, for valuable comments and Chong Wang for identifying shortcomings of the effective field theory. S.G. acknowledges support from the Israel Science Foundation (ISF) Grant no. 586/22 and the US–Israel Binational Science Foundation (BSF) Grant no. 2020264. S.M. is supported by Vetenskapsr\aa det (grant number 2021-03685), Nordita and STINT. U.B. acknowledges support from the Israel Academy of Sciences and Humanities through the Excellence Fellowship for International Postdoctoral Researchers.

\widetext
\pagebreak
\begin{center}
\textbf{\large 
\large Supplemental Material: Deconfined quantum criticality in Ising gauge theory entangled with single-component fermions}
\end{center}
\setcounter{equation}{0}
\setcounter{figure}{0}
\setcounter{table}{0}
\setcounter{page}{1}
\makeatletter
\renewcommand{\theequation}{S\arabic{equation}}
\renewcommand{\thefigure}{S\arabic{figure}}
\renewcommand{\bibnumfmt}[1]{[S#1]}

\section{Dimerization from short-range density-density interactions}
\begin{figure}[h]
    \captionsetup[subfigure]{labelformat=empty}
    \subfloat[\label{subfig:U_configs}]{}
    \subfloat[\label{subfig:U_region}]{}
    \vspace{-10pt}
	\includegraphics[width=0.7\linewidth]{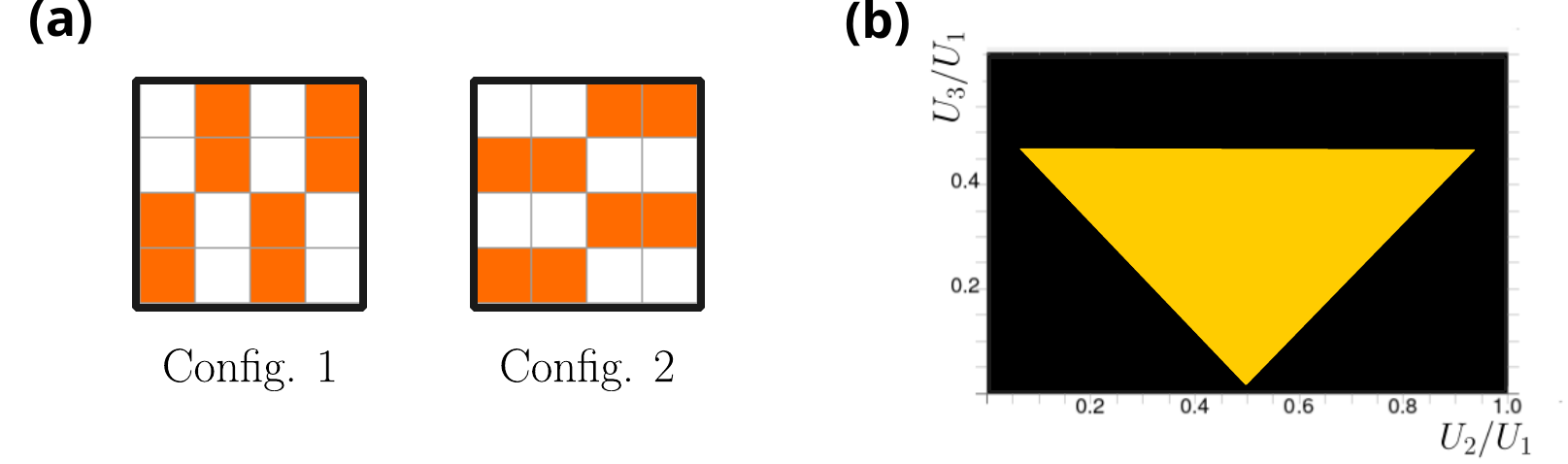}
	\caption{(a) Example of configurations of classical dimers which reproduce the pattern observed in the large-$h$ limit of the gauge theory at half filling. (b) The yellow region in the $U_2/U_1$-$U_3/U_1$ plane is where this dimerized configuration has the minimal classical energy among all possible configurations of eight particles on a $4\times 4$ grid. In our numerical calculations we choose $U_2/U_1=0.5$ and $U_3/U_1=0.25$, which lies exactly in the middle of the highlighted region. This choice implies maximal stability of this particular dimerization pattern to fluctuations.}
\label{supfig:U_plot}
\end{figure}
As argued in the main text, it is desirable to find a way to realize the translation symmetry breaking pattern induced by the $\Zt$ electric field even in the absence of tension in gauge electric strings. To this end, we introduce a suitable combination of short-range NN, NNN and NNNN density-density interactions. To stabilize the proper pattern, we first find values $U_1$, $U_2$, $U_3$ for which the classical Hamiltonian
\begin{equation}
    H_U=U_1\sum_{\langle \rr,\rr'\rangle}n_\rr n_{\rr'}+U_2\sum_{\langle \langle \rr,\rr' \rangle \rangle}n_\rr n_{\rr'}+U_3\sum_{\langle \langle \langle \rr,\rr' \rangle \rangle \rangle}n_\rr n_{\rr'}
    \label{eq:H_u_sup}
\end{equation}
has the D-CDW* configuration (shown in Fig. \ref{subfig:U_configs}) as a ground state. To determine these values, we evaluate the energy function \eqref{eq:H_u_sup} of every possible arrangement of eight fermions on a $4\times 4$ torus, corresponding to half filling. Despite the fact that the size for the ground state unit cell is only $2\times 4$, this particular pattern requires non-vanishing NNNN interactions. We find that the energy of the desired configuration is
\begin{equation}
    E_{\text{CDW}}(U_1,U_2,U_3)=4 U_1+8 U_2+4 U_3,
    \label{eq:E:d_cdw}
\end{equation}
while all the other arrangements lead to a total of $54$ different energies. We adopt a heuristic approach, and by direct inspection we find that in order for \eqref{eq:E:d_cdw} to be an energy minimum all $U_k$ need to be positive, i.e. the interactions are repulsive. After narrowing down the number of possible choices of the coefficients, we are able to determine that the values given in the main text indeed stabilize the desired configuration. This corresponds to $U_1=U$, $U_2=U/2$ and $U_3=U/4$, so that there is a single parameter controlling the overall strength of the interactions. It is important to understand how fine tuned these values are. Indeed, the presence of another configuration very close in energy to the D-CDW* could cause numerical issues once the energy of hopping fermions is taken into account. By performing a scan in the $U_2/U_1$-$U_3/U_1$ plane, we find that the configuration represents an energy minimum within a triangular region (see Fig. \ref{subfig:U_region}) centered at the selected values of $U_k$, which are therefore the best choice available. The size of the gap is determined by $U_3$, which needs to be chosen large enough to avoid instabilities. 

\section{Stabilizing a $\pi$ flux in the deconfined phase}
While a negative magnetic coupling $J$ obviously stabilizes the $\pi$ flux, it is not necessary at small $U$ and $h$ as a negative magnetic coupling is generated perturbatively even at $J=0$ by the hopping of the itinerant fermions. This is a manifestation of Lieb's theorem \cite{lieb1994flux}, which states under very general conditions that the lowest energy of a fermionic system at half filling is realized in the $\pi$ flux phase. 

While the perturbative argument still holds in the presence of a weak repulsive interaction $U$, it becomes ineffective for large values of $U$, as deep in the dimerized phase the fermions are frozen and the hopping processes are inhibited. In this case, in the absence of magnetic coupling $J$, we expect confinement to occur already for small values of $h$. Since we want to observe the splitting of the transition, we need  instead to have an extended stable $\pi$ flux deconfined phase even at large values of $U$. To this end, we introduce a sizable negative magnetic coupling $J$ at large $U$ according to the equation $J=-U/40$. The small value of the prefactor is due to the fact that the SSB transition is actually stabilized by the NNN-nearest-neighbor interaction of strength $U_3=U/4$, hence we need to tune to relatively large values of $U/h$.   

\section{The Fredenhagen-Marcu order parameters}
\begin{figure}[t]
    \captionsetup[subfigure]{labelformat=empty}
    \subfloat[\label{subfig:fm_e}]{}
    \subfloat[\label{subfig:fm_m}]{}
    \subfloat[\label{subfig:fm_e_plot}]{}
    \hspace{-10pt}
	\includegraphics[width=0.8\linewidth]{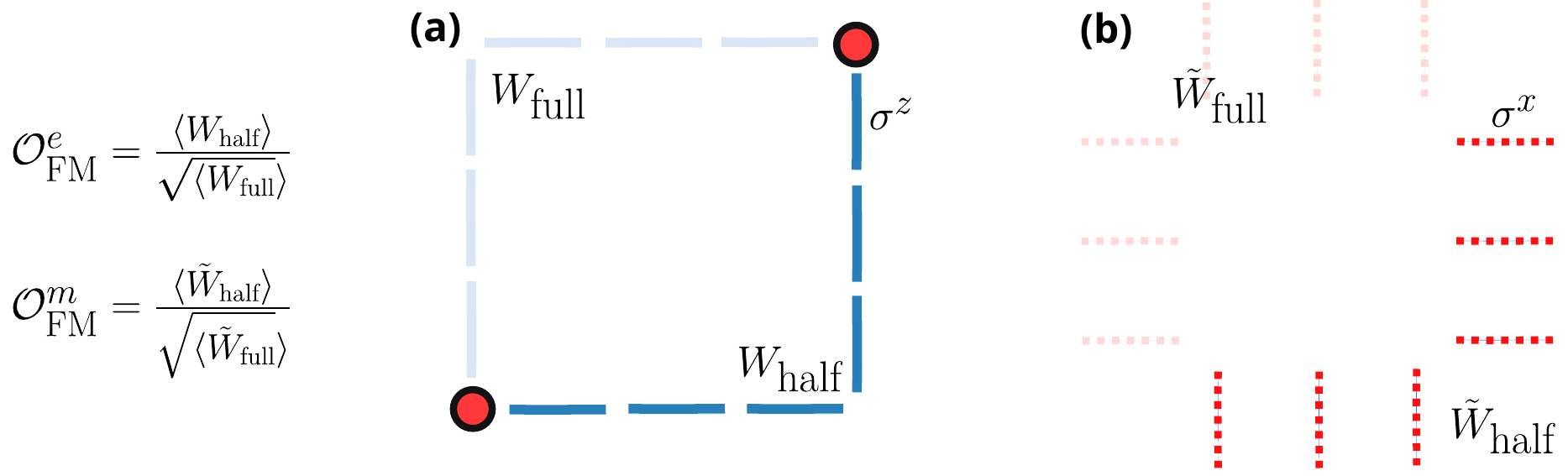}
	\caption{Definitions of the electric (a) and magnetic (b) Fredenhagen-Marcu order parameters. The open strings detect condensation of $e$ and $m$ particles respectively, while the denominator provides the correct normalization.}
\label{supfig:fms}
\end{figure}
In the presence of matter it is well known that the Wilson loop cannot be used to detect confinement of charges, since it follows a perimeter law both in the confined and in the deconfined phase as a consequence of charge screening. A suitable generalization, the so called Fredenhagen-Marcu order parameter \cite{Fredenhagen_1986}, exists and has been employed successfully to diagnose confinement in a context similar to ours \cite{Gregor_2011, Verresen_2021}. The idea is to calculate the expectation value of a half-loop with gauge invariant endpoints, and normalize it by dividing it by the \textit{square root} of the expectation value of the full loop. With this definition, the order parameter has a finite expectation value in the Higgs-confined phase and vanishes in the deconfined phase. As shown in Fig. \ref{supfig:fms}, two such loops can be defined. The  ``electric'' one (Fig. \ref{subfig:fm_e}) is constructed from the Wilson line, of which it is the proper generalization. Since in this case a half loop is not gauge-invariant, it must terminate with a $\Zt$ charged endpoint which in our case is a fermionic $e$-particle. We refer to this operator by $\mathcal{O}_{\text{FM}}^e$. By analogy, a magnetic order parameter $\mathcal{O}_{\text{FM}}^m$ is defined by applying the same construction to the t'Hooft loop formed by $\sigma^x$ operators, see Fig. \ref{subfig:fm_m}. In this case, the half-loop is gauge-invariant by itself, with its endpoints corresponding to $m$-particles (visons). 

The physical meaning of these order parameters is that they detect condensation of $e$ and $m$ particles, which corresponds to confinement of $m$ and $e$, respectively. While away from the special lines $t=0$ and $h=0$ both order parameters are in principle useful, for large values of $U$ the electric order parameter $\mathcal{O}_{\text{FM}}^e$ takes small values in both phases, making it more difficult to be a useful numeric diagnostics in that regime. To detect confinement of gauge charges,  we use the magnetic order parameter $\mathcal{O}_{\text{FM}}^m$, as shown in Fig. 2 of the main text. The drawback is that on a cylinder of circumference $L_y=4$, that was used by us, only a small loop that encloses a $2\times 2$ square of plaquettes fits properly. Therefore for $L_y=4$, we cannot check how this order parameter sharpens as its perimeter is increased. 

\section{Large $U$ regime: confinement in the background of dimerized fermions}

\begin{figure}[h]
    \hspace{-10pt}
    \includegraphics[width=0.8\linewidth]{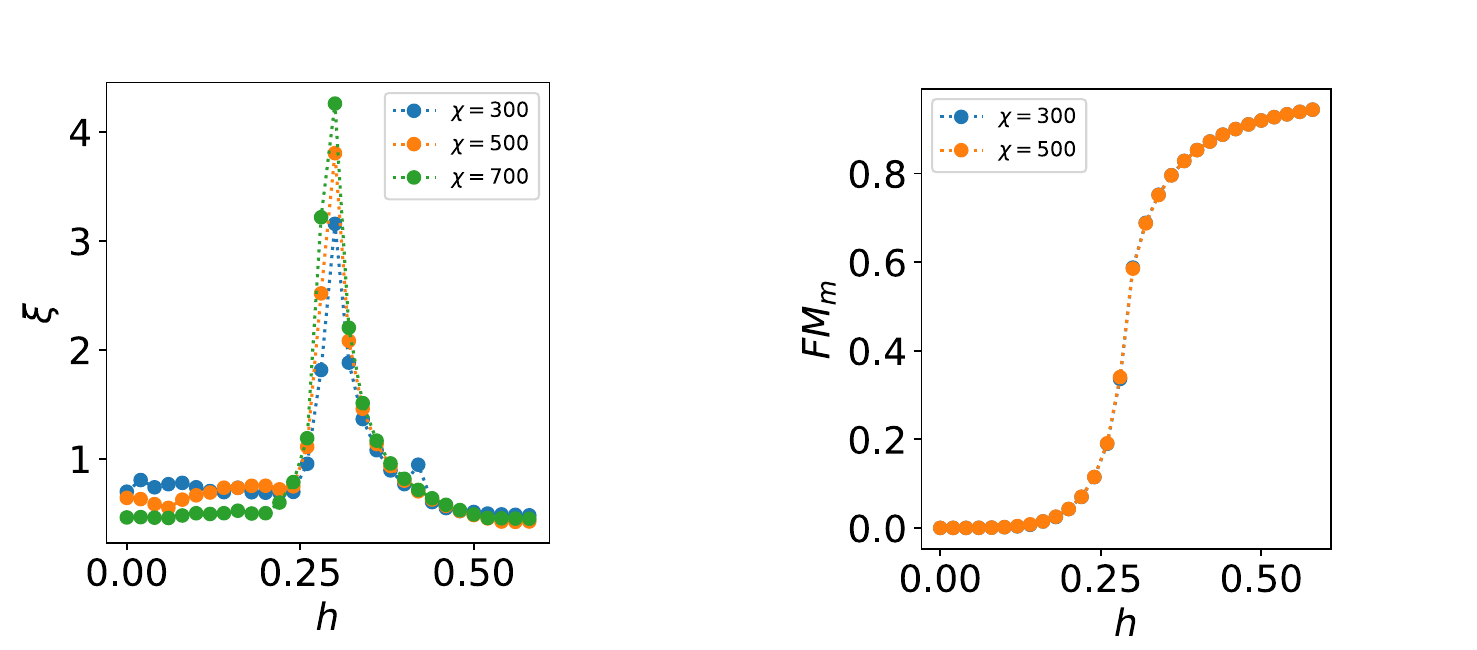}
	\caption{Correlation length (left) and FM order parameter (right) as a function of the electric coupling at large $U/t=20$. The order parameter rises in correspondence of the peak in $\xi$. We provide data for different values of the bond dimension $\chi$. The correlation length diverges at the peak and, in general, converges slowly. A semi-local observable like the FM order parameter, on the other hand, converges to the desired accuracy for moderate values of $\chi$ even near the confinement transition.}
 \label{supfig:large_U}
\end{figure}

As argued in the main text, at large $U$ the fermions are frozen in place and we predict a pure confinement transition in the gauge sector. In Fig. \ref{supfig:large_U} we provide additional numerical results which highlight the transition.

In the following, we provide details on our analysis that indicates that the confinement transition in the Ising gauge theory half-filled with fermions is of the Ising* universality class in the large-$U$ regime. 

\begin{figure}[h]
    \hspace{-10pt}
    \includegraphics[width=0.6\linewidth]{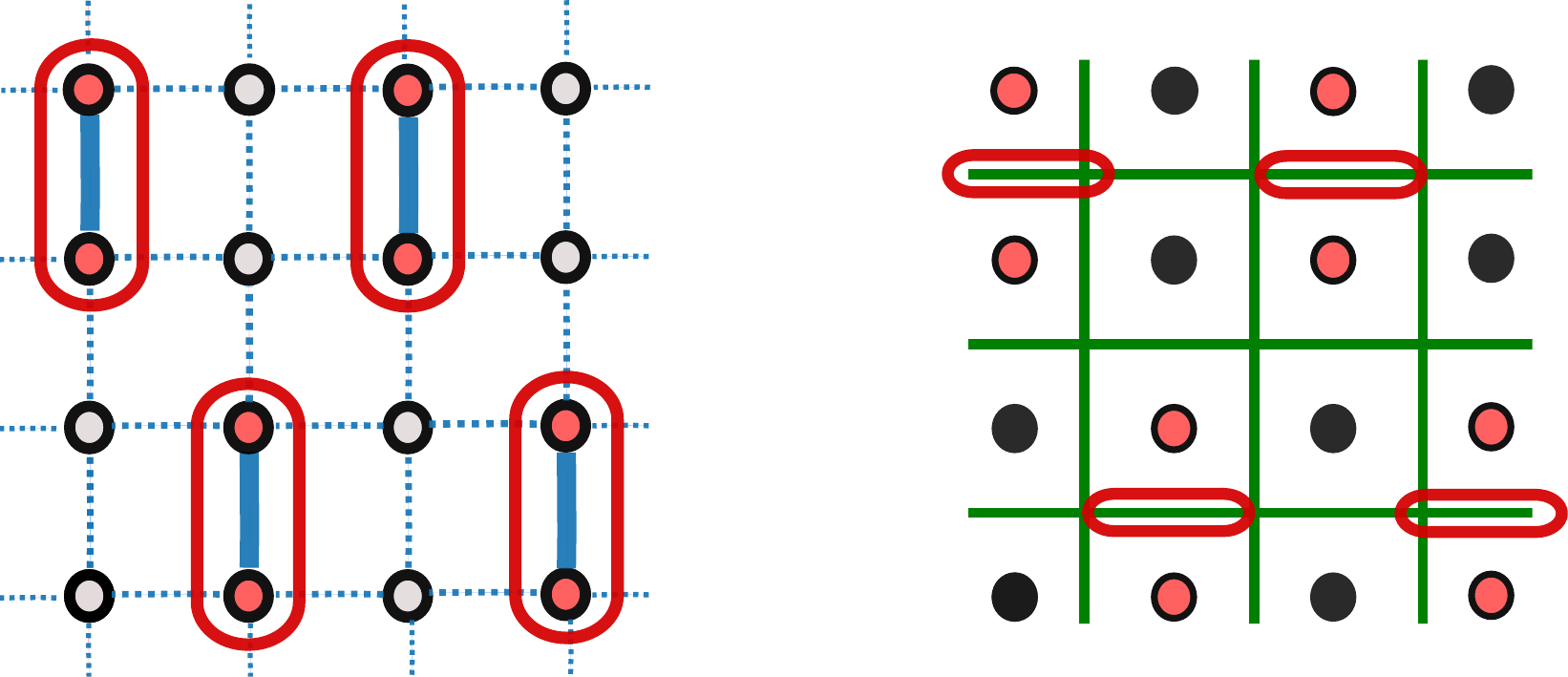}
	\caption{Left: Translation symmetry breaking pattern of fermions frozen on the original square lattice. Right: The half-frustrated Ising model is defined on the (green) sites of the dual lattice. In a particular gauge (chosen here) the frustrated Ising bonds are highlighted in red.}
 \label{supfig:SSB_1}
\end{figure}

\subsection{Duality: half-frustrated Ising model}
We assume that the fermions are frozen into the pattern indicated in Fig. \ref{supfig:SSB_1} (left). As a result, we are dealing with a pure Ising gauge theory with an alternating Gauss law $G_{\mathbf{r}}=\pm 1$. We can map that to a half-frustrated Ising model following the non-local duality due to Wegner \cite{wegner1971duality}. To this end, we first define Pauli matrices acting on the dual lattice
\begin{equation} \label{Wd}
    X_\mathbf{r}*= \prod_{b\in \square_\mathbf{r}*} \sigma^z_b, \qquad
    Z_\mathbf{r}*= \prod_{b\in \text{vertical string}} \sigma_b^x,
\end{equation}
where the semi-infinite vertical string emanates from the dual site $\mathbf{r}*$. 
Using this duality together with the Gauss law, one can rewrite the Ising gauge theory Hamiltonian as
\begin{equation}
    H=-J\sum_{\mathbf{r}*} X_{\mathbf{r}*} - h \sum_{\langle \mathbf{r}*, \mathbf{r'}*\rangle} s_{\langle \mathbf{r}*, \mathbf{r'}*\rangle} Z_\mathbf{r}* Z_\mathbf{r'}*,
\end{equation}
where the frustration factor $s_{\langle\mathbf{r}*, \mathbf{r'}*\rangle}=\pm 1$ follows a simple pattern illustrated in Fig. \ref{supfig:SSB_1} (right). Independent of how the semi-infinite string in Eq. \eqref{Wd} is defined, one finds that the products of the Ising couplings on dual plaquettes encircling  static fermions must be negative. So we are dealing with a version of the frustrated Ising model, where only half of the dual plaquettes are frustrated. 

\subsection{Vison band structure}
\begin{figure}[t]
    \hspace{-10pt}
    \captionsetup[subfigure]{labelformat=empty}
    \subfloat[\label{subfig:dual_lat}]{}
    \subfloat[\label{subfig:unit_cell}]{}
    \subfloat[\label{subfig:density_plot}]{}
     \includegraphics[width=0.9\linewidth]{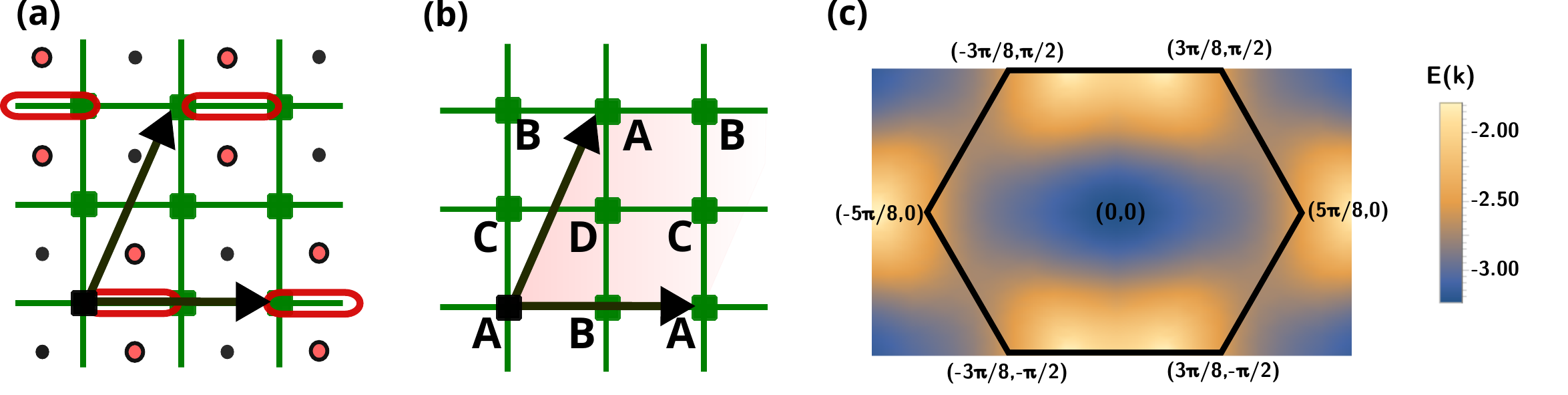}
	\caption{(a): Visons hopping on the dual square lattice (green) see frozen fermions (red) as $\pi$-fluxes. (b): Oblique Bravais lattice with a four-site unit cell. (c): Density plot of the vison dispersion, highlighting the presence of single minimum at $(k_x,k_y)=(0,0)$. The hexagonal Brillouin zone corresponding to the real-space periodicity of the system is drawn in black.}
\end{figure}

Here we determine the band structure \footnote{In $\mathbb{Z}_2$ gauge theory the number of visons is conserved only modulo 2. If one embeds the Ising gauge theory into a parent $u(1)$ gauge theory and neglects all monopole contributions, the vison number is conserved.} of visons that live on the dual square lattice. We assume that fermions are frozen into the dimerezed translation-breaking pattern. A vison collects a $\pi$ phase when moved around a fermion. In our calculation this is implemented by swapping the sign of the vison hopping parameter on links that are adjacent to the pairs of frozen fermions, see Fig. \ref{subfig:dual_lat}. As a result, we end up with a four-site unit cell repeated in an oblique Bravais lattice with the primitive translation vectors $\mathbf{a}_1=(2,0)$ and $\mathbf{a}_2=(1,2)$, see Fig. \ref{subfig:unit_cell}.   

The resulting four-band tight-binding vison Hamiltonian in momentum space can be obtained by assigning hopping phases between sites, labeled as in \cref{subfig:unit_cell}, within and out of the unit cell. Sites in the same unit cell are assigned zero relative displacement, while relative displacements between different unit cells are given by the primitive vectors $\mathbf{a}_1$ and $\mathbf{a}_2$. In momentum space one gets
\begin{equation}
    H_v=-t_v
\left(
\begin{array}{cccc}
 0 & 1+e^{-2 i k_x} & 1 & e^{-i (k_x+2 k_y)} \\
 1+e^{2 i k_x} & 0 & e^{i (k_x-2 k_y)} & 1 \\
 1 & e^{-i (k_x-2 k_y)} & 0 & 1-e^{-2 i k_x} \\
 e^{i (k_x+2 k_y)} & 1 & 1-e^{2 i k_x} & 0 \\
\end{array}
\right).
\end{equation}

By examining the resulting band structure, we observe that the global energy minimum is unique and located at $k_x=k_y=0$, see fig \ref{subfig:density_plot}. Qualitatively, it is different to the case fully filled with static charges (odd $\mathbb{Z}_2$ gauge theory), where the vison dispersion has a two minima degenerate in energy \cite{sachdev2018topological}. In fact, the result here is qualitatively similar to what one finds for visons in the absence of any matter (even $\mathbb{Z}_2$ gauge theory). Therefore it is expected that the phase transition is in the Ising* universality class.

\section{Translation symmetry breaking transition at $h=0$}
Here we briefly discuss a scenario for the translation symmetry breaking quantum criticality driven by the short-range repulsion interactions at vanishing string tension ($h=0$). At $h=0$, gauge fluctuations are absent. Due to Lieb's theorem \cite{lieb1994flux}, on the square lattice a background homogeneous $\pi$-flux is stabilized in the ground state. The zigzag dimerized pattern, preferred by strong short-range interactions, leads to a four-site oblique unit cell, with primitive translation vectors $\mathbf{a}_1=(2,0)$ and $\mathbf{a}_2=(1,2)$, see Fig. \ref{supfig:zz} (left). The reciprocal lattice is generated by the vectors $\mathbf{b}_1=2\pi(1/2,-1/4)$ and $\mathbf{b}_2=2\pi(0,1/2)$. As a result, the folded Brillouin zone has a form of an irregular hexagon illustrated in Fig. \ref{supfig:zz} (right). 

The resulting four-band tight-binding fermionic kinetic Hamiltonian in the background $\pi$-flux in momentum space is
\begin{equation}
    H=-t
    \left(
\begin{array}{cccc}
0 & -1-e^{-2 i k_x} & 1 & e^{i (-k_x-2 k_y)} \\
-1-e^{2 i k_x} & 0 & e^{i (k_x-2 k_y)} & 1 \\
1 & e^{i (2 k_y-k_x)} & 0 & 1+e^{-2 i k_x} \\
e^{i (k_x+2 k_y)} & 1 & 1+e^{2 i k_x} & 0 \\
\end{array}
\right).
\label{eq:tight_binding_dirac}
\end{equation}
This results in two Dirac points which, within the Brillouin zone, are at momenta $(\pm \pi/2, 0)$. 

\begin{figure}[ht]
    \hspace{-10pt}
	\includegraphics[width=0.6\linewidth]{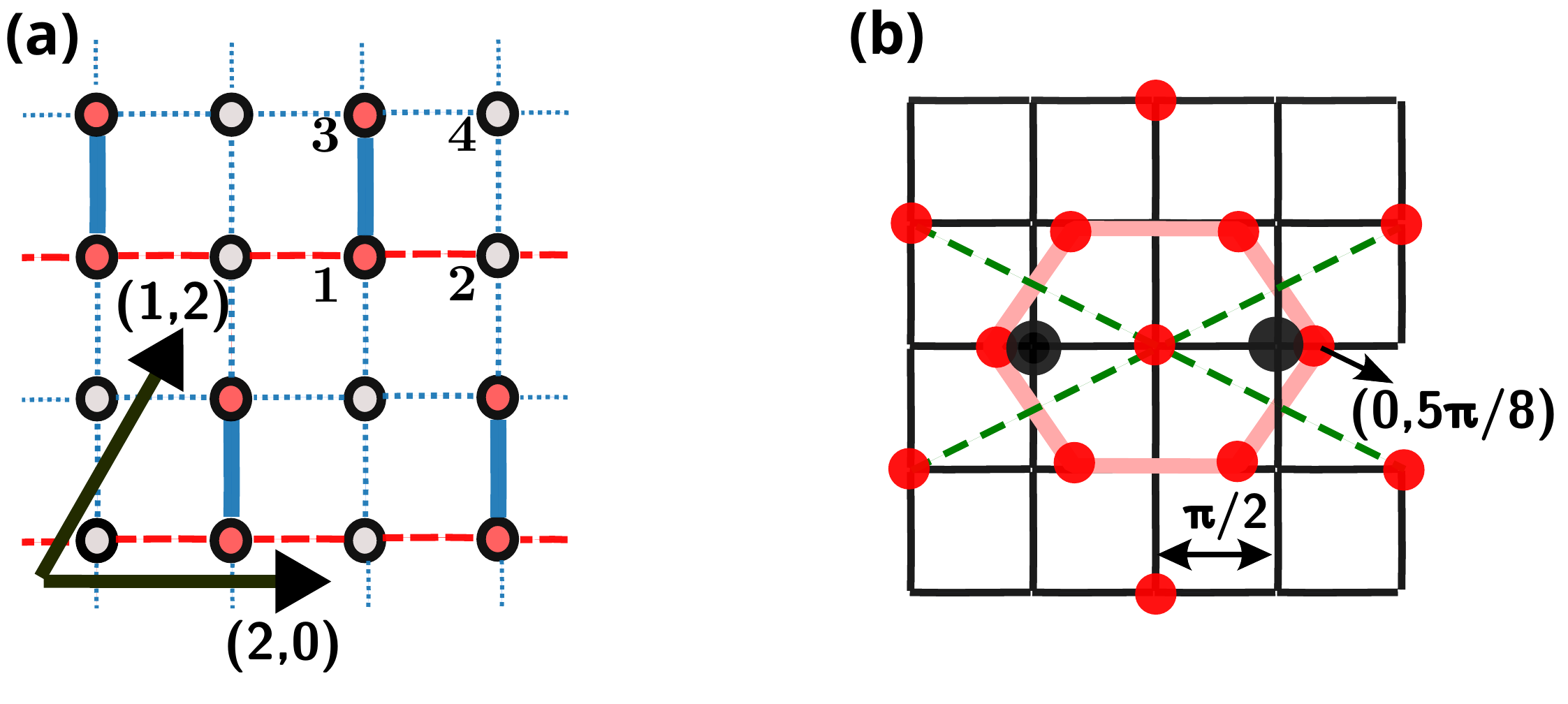}
	\caption{Left: The zig-zag symmetry breaking pattern implies a four-site unit cell. In our calculation, the background $\pi$-flux is implemented with horizontal frustrated hoppings along red solid lines. Green numbers denote our choice of ordering of bands. Right: Folded Brillouin zone is displayed in red. The two Dirac points (black dots) are at momenta $(\pm \pi/2, 0)$.}
 \label{supfig:zz}
\end{figure}

Given that within our choice of the ordering of sites (shown in Fig. \ref{supfig:zz} (left)) in the enlarged unit cell, the symmetry-breaking density pattern is $\delta n=(+,-,+,-)$, the relevant order parameter $\rho$ couples to the fermionic sector as

\begin{equation}
    \rho\psi^\dagger 
    \left(
\begin{array}{cccc}
 1 & 0 & 0 & 0 \\
 0 & -1 & 0 & 0 \\
 0 & 0 & 1 & 0 \\
 0 & 0 & 0 & -1 \\
\end{array}
\right)
\psi.
\label{eq:op_coupling}
\end{equation}

By direct calculation of the resulting energy bands, we find that the term \eqref{eq:op_coupling} gaps out the Dirac cones.

\begin{figure}[ht]
    \hspace{-10pt}
	\includegraphics[width=0.8\linewidth]{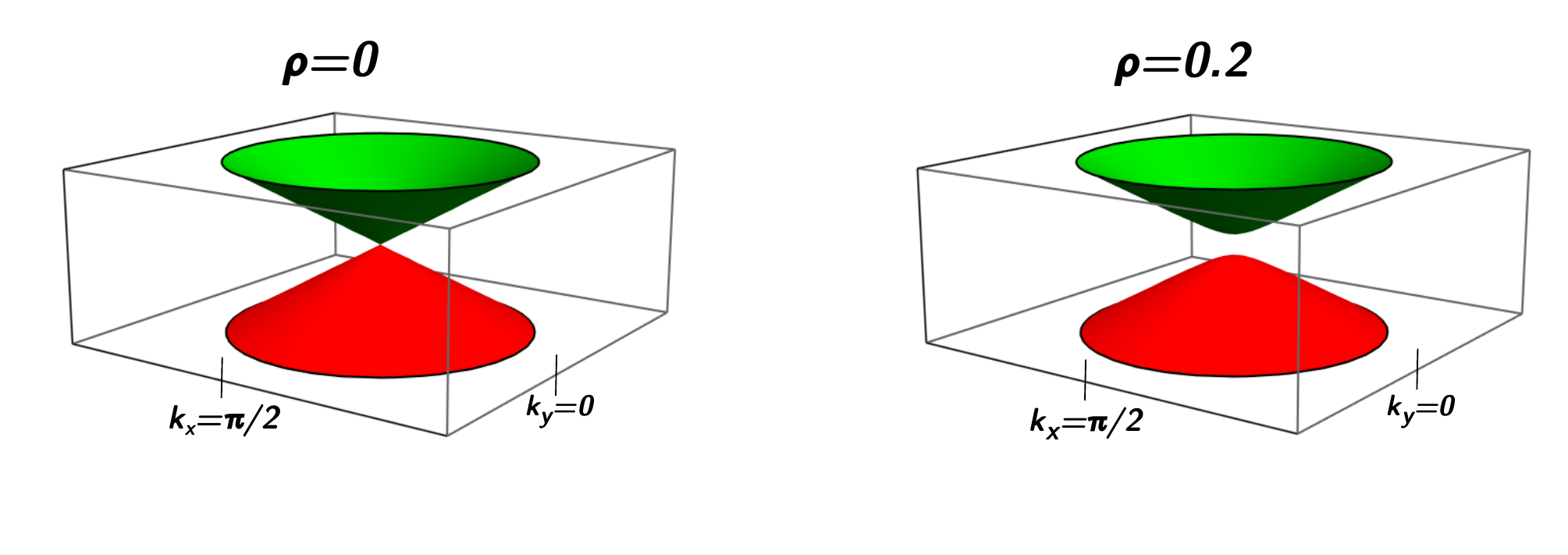}
	\caption{Energy bands of the tight binding Hamiltonian \eqref{eq:tight_binding_dirac}, zoomed in around one of the two Dirac cones, for order parameter $\rho=0$ (left) and $\rho=0.2$ (right). One can see that a finite order parameter acts as a mass term, gapping out the Dirac cone.}
 \label{supfig:zz}
\end{figure}

\section{Mapping to a gauge-invariant spin model}
In Ref. \cite{borla2022}, see also \cite{PhysRevResearch.5.023077, PhysRevD.108.074503}, it is explained how models with spinless fermions hopping on a square lattice and coupled to $\Zt$ gauge fields can be rewritten in a gauge-invariant form built of spin $1/2$ degrees of freedom living on the links of the lattice. The mapping is exact, and allows to express all the possible observables and Hamiltonians in terms of gauge-invariant Pauli matrices. From the computational point of view, the advantage of this approach is two-fold: (1) the spin model is devoid of any gauge redundancy, consisting of link degrees of freedom only, and (2) the Gauss law is encoded exactly and does not need to be enforced energetically, which would introduce an additional convergence parameter into the simulations.

First, we introduce the Majorana operators 
\begin{equation}
\gamma_\rr = c^{\dagger}_\rr+c_\rr^{\vphantom{\dagger}},\,\,\,\,\,\,\,\,\,\,\,\,
\tilde{\gamma}_\rr = i(c^{\dagger}_\rr-c_\rr^{\vphantom{\dagger}}).
\end{equation}
Using now the $\mathbb{Z}_2$ gauge fields and Majorana variables, we construct gauge-invariant Pauli operators 
\begin{equation}
\begin{aligned}
X_{\rr,\eta}&=\sigma^x_{\rr,\eta},\\
Z_{\rr,\hat{x}} &= -i\tilde{\gamma}_\rr \sigma^z_{\rr,\hat{x}}\gamma_{\rr+\hat{x}}\sigma^x_{\rr+\hat{x},-\hat{y}},\\
Z_{\rr,\hat{y}} &= -i\tilde{\gamma}_\rr \sigma^z_{\rr,\hat{y}}\gamma_{\rr+\hat{y}}\sigma^x_{\rr,\hat{x}},
\label{eq:mapping}
\end{aligned} 
\end{equation}
where in $Z_{\rr,\eta}$ the $\sigma^x$ factors are chosen so that 
the appropriate spin commutation relations are satisfied between the gauge-invariant variable on all links. 

\begin{figure}[ht]
    \hspace{-10pt}
	\includegraphics[width=0.8\linewidth]{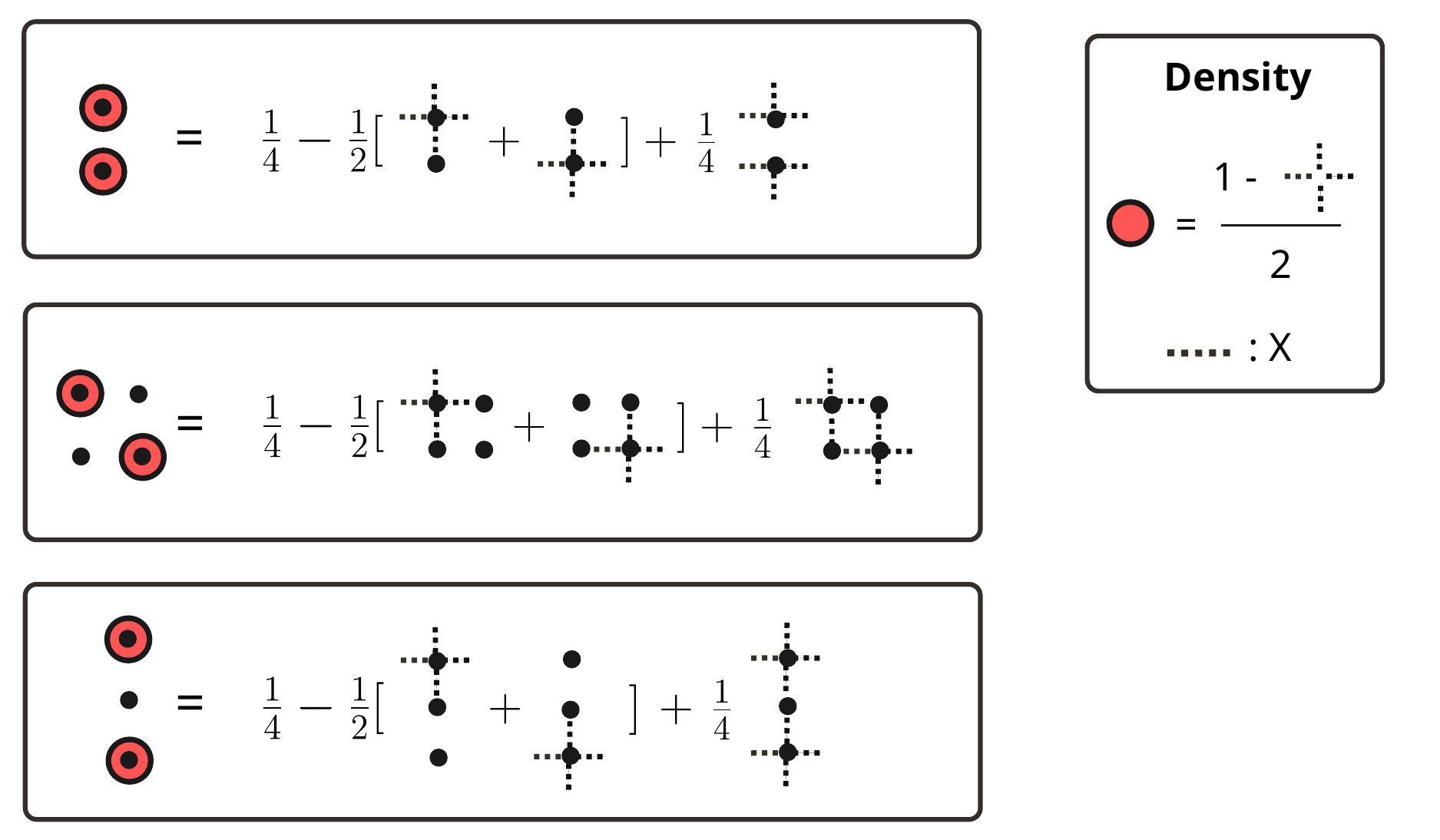}
	\caption{Graphical representation of the expressions for the NN, NNN and NNNN density-density interactions after the mapping \eqref{eq:mapping}. These are obtained by taking appropriate products of number operators (red blobs) from Eq. \eqref{eq:n_map_2d} on the relevant sites.}
 \label{supfig:nnn}
\end{figure}

By combining the above definitions with the Gauss law constraint, which reads
\begin{equation}
i \tilde{\gamma}_\rr\gamma_\rr=\prod_{b \in +_{\rr}}\sigma^x_b,
\label{eq:gauss_majorana}
\end{equation}
one can map each term in the Hamiltonian, see Ref. \cite{borla2022} for details.  
Although the short-range $U$ interactions were not considered there, this can be fixed in a straightforward manner.  Namely, any term that depends on the local fermionic densities only can be easily mapped using the Gauss law exclusively.
In particular, 
\begin{equation}
n_{\rr} = \frac{1-\prod_{b\in +_\mathbf{r}} X_b}{2},
\label{eq:n_map_2d}
\end{equation} 
and so up to a constant the number operator on a site maps onto a star operator emanating from that site. The nearest-neighbor, next-nearest-neighbor and next-next-nearest-neighbor interactions introduced in the main text are mapped by taking appropriate products of \eqref{eq:n_map_2d}. The results are summarized in Fig. \ref{supfig:nnn}.

\section{Numerical methods}
As the main numerical tool, we use the infinite density matrix renormalization group (iDMRG) algorithm to obtain ground states in the form of infinite matrix product states (iMPS). All our simulations are performed with the Python tensor network package TeNPy \cite{hauschild2018efficient}. The iDMRG algorithm is designed to study one-dimensional systems, but it can be adapted to two dimensions. While for systems which are finite in one of the two directions (in this paper $y$, which is taken to be periodic) this is easily done by snaking, i.e., ordering the lattice sites and considering the corresponding 1d system, the procedure introduces artificial long-range interactions that severely limit the performance of the algorithm. The entanglement of such systems grows exponentially with the size of the circumference, which makes reaching the thermodynamic limit by finite size scaling challenging.  Since we require large accuracy to determine whether the confinement and SSB transitions occur together at the exotic quantum critical point, in this work we limit our simulations to circumference sizes of $L_y=4$. 

\subsubsection{Details on the convergence}
\begin{figure}[t]
    \hspace{-10pt}
    \captionsetup[subfigure]{labelformat=empty}
    \subfloat[\label{subfig:conv_E}]{}
    \subfloat[\label{subfig:conv_trunc}]{}
     \includegraphics[width=0.9\linewidth]{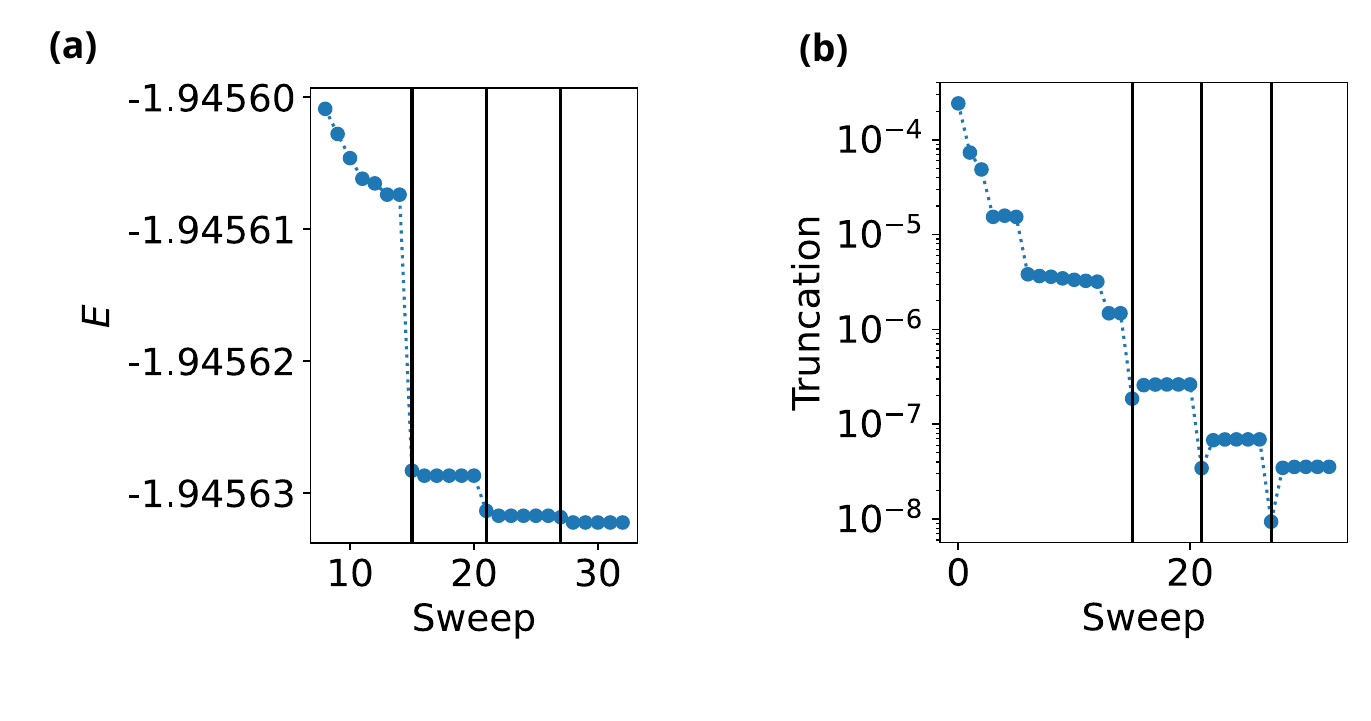}
	\caption{We display the values of the energy (left) and of the minimum Schmidt coefficient determining the truncation in the iDMRG algorithm (right) as a function of the iteration. The data correspond to the point in parameter space $U/t=2$, $h/t=0.65$. The vertical lines denote the sweeps where we increase the bond dimension $\chi$, up to the maximum value $\chi=1000$. }
 \label{supfig:conv}
\end{figure}

The convergence of the algorithm is achieved once the energy difference $\Delta E$ between two successive iterations is smaller than a certain threshold, which we typically set to $10^{-7}$. We also perform an additional check on the entanglement entropy, making sure that the difference $\Delta S$ becomes smaller than $10^{-3}$. To make sure that the minimization has worked properly we look at a number of indicators, such as the energy and the smallest Schmidt coefficient, and check that they decrease monotonically as a function of the iteration. As an example, shown in Fig. \ref{supfig:conv}, we consider the point in parameter space $U/t=2$, $h/t=0.65$, close to the exotic phase transition where the convergence of the algorithm is potentially more problematic. We start the algorithm with a moderate bond dimension $\chi=250$ and we run it until convergence is achieved. Then, starting with the iMPS wave function that we obtain, we further optimize it by increasing the bond dimension and let the algorithm converge to a new minimum of the energy. We repeat the procedure up to the maximum bond dimension $\chi=1000$.  

Such conditions are strict enough that the convergence of local observables is in practice always guaranteed even in the vicinity of gapless points. The entanglement entropy and the correlation length, on the other hand, converge slower and are expected to diverge as a function of the bond dimension whenever the system is gapless. 

\subsubsection*{$U(1)$ conservation and fermion filling}
In simulating lattice gauge theories in their original formulations, one faces the hurdle of implementing the gauge constraint. This can be done energetically, by adding a large term to the Hamiltonian that penalizes the states which do not satisfy the Gauss law. While this works, it introduces a numerical error and one extra convergence parameter. The mapping introduced in Ref \cite{borla2022} and reviewed in the section above, however, allows to implement the constraint exactly. Using the dual spin system has one drawback: current TeNPy iDMRG implementations do not allow to conserve multi-link charges such as the $U(1)$ number operator $N=\sum_{\mathbf{r}}n_{\mathbf{r}}=\sum_{\mathbf{r}} \frac{1-\prod_{b\in +_\mathbf{r}} X_b}{2}$, where the last equality follows from Eq. \eqref{eq:n_map_2d}. Therefore, in this formulation, the fermion filling must be tuned by changing the chemical potential $\mu$. 
It was found in \cite{borla2022, PhysRevB.105.245105} that for the vanishing magnetic coupling $J=0$ the half-filling is achieved by setting $\mu=h$.

\subsubsection*{Correlation length}
Once the ground state is obtained in the form of an infinite matrix product state (iMPS), one can calculate the correlation length $\xi$. To this end, similar to statistical mechanics, one can construct a transfer matrix and examine its eigenvalues. As explained in detail in \cite{hauschild2018efficient}, the behavior of the slowest decaying correlation function can be inferred from the second largest eigenvalue of the MPS transfer matrix, which is readily computed in our simulations. In gapless systems quantum fluctuations occur over all length scales, and the correlation length is expected to diverge. This is particularly useful to detect a quantum critical point between gapped phases, corresponding to second-order phase transitions. Peaks in the correlation length, which increase as a function of the bond dimension $\chi$, are a clear signature of quantum criticality that can be obtained without knowing further details about the system.
\end{document}